\documentclass[11pt]{article}

\usepackage[margin=1in]{geometry}
\usepackage{amsmath, amssymb, amsthm, mathtools}
\usepackage{xcolor}
\usepackage{booktabs}
\usepackage{graphicx}
\usepackage{float}
\usepackage{tikz}
\usetikzlibrary{arrows.meta, positioning, decorations.pathreplacing, calc}
\usepackage[colorlinks=true, linkcolor=blue, citecolor=blue, urlcolor=blue]{hyperref}
\usepackage[round]{natbib}

\newcommand{\E}{\mathbb{E}}

\newtheorem{theorem}{Theorem}
\newtheorem{lemma}{Lemma}
\theoremstyle{remark}
\newtheorem{remark}{Remark}

\title{Fold Weighting for Cross-Validation\\ Under Continuous Distributional Drift}
\author{Yegor Baranovski\\ \href{mailto:yegorbar01@gmail.com}{\texttt{yegorbar01@gmail.com}}}
\date{}

\begin{document}
\maketitle

\begin{abstract}
Cross-validation weights all folds equally, so under drift the oldest folds count as much as the newest. We model drift by a H\"older condition in total variation distance and give finite-sample bounds on the excess deployment risk of the configuration that sliding-window cross-validation selects. Each bound splits into drift, estimation, and validation-noise terms. With equal weights the drift terms saturate as the record grows. With fold weights fixed in advance, the drift terms become weighted averages, which shrink as weight moves toward deployment. The noise term grows as the weights concentrate and shrinks with the batch size. The drift terms are linear in the weights and the noise term is convex, so minimizing the bound over the weights is a convex problem. Applications to U.S. earnings and flight delays show both sides of this trade-off. Weight on recent folds pays off when a change in the best predictor persists and each period has many observations, and equal weights do best when changes are short-lived.
\end{abstract}

\section{Introduction}\label{sec:intro}

A prediction model is trained on the past and deployed in the future. In many real-world applications,
the data-generating distribution can move in between. Climate patterns, 
animal populations, and credit conditions change gradually, not only through discrete breaks.

Cross-validation is the standard tool for choosing a model, and its classical justification assumes 
exchangeable data. Under drift it estimates risk under a mixture of past distributions, 
often not under the distribution the model will face. The usual remedy is sliding-window cross-validation. Each fold fits on a recent window and validates on the periods that follow. Theory for this practice asks how much history to use, by choosing a window width or which data enters a comparison \citep{han2024model, mazzetto2023adaptive, capponi2025nonstationarity}. Once the window is fixed, the folds still enter the criterion with equal weight, however far each sits from deployment.

We ask how the fold weights affect the configuration that cross-validation selects. We model drift by a H\"older condition on the path of distributions in total variation distance, and give two finite-sample bounds on the excess deployment risk of the configuration that sliding-window cross-validation selects. Both split into drift, estimation, and validation-noise terms.
\begin{enumerate}
    \item With equal weights (Theorem~\ref{thm:upper_bound}), the drift terms average over all folds, so the earliest folds count as much as the latest, and the bound saturates at its trivial value as the record grows.
    \item With fold weights fixed in advance (Theorem~\ref{thm:weighted_bound}), 
    the drift terms become weighted averages, so weight near deployment shrinks them. The noise term pays for this, since equal weights over $K$ folds minimize it, and putting all weight on the latest fold multiplies it by $\sqrt{K}$.
\end{enumerate}
In a simulation (Section~\ref{sec:sim}), equal-weight cross-validation picks the worst of four configurations once the record is long, and weights that decay with the age of the fold do not. On fifty years of U.S. earnings (Section~\ref{sec:cps}), where the best single predictor of pay changes once, from sex to education in 1983, equal weights lag the change, and the last fold alone is worst when samples are small and best when they are large. On sixteen years of U.S. flight delays (Section~\ref{sec:flights}), where the best predictor changes with the seasons and the changes do not last, the last fold chases them and, from 250 flights per month up, equal weights are best. A decay rate fixed in advance is never the worst of the three. Favoring recent folds is right when distributions drift further apart with time; when drift recurs, the right weights depend on the latent path of distributions, which must itself be known or estimated (Section~\ref{sec:future}).
Section~\ref{sec:setup} sets up the drift model and the target, Section~\ref{sec:cv} the procedure, Section~\ref{sec:results} the results, Section~\ref{sec:sim} a simulation, Section~\ref{sec:apps} applications to earnings and flight delays, and Section~\ref{sec:future} open problems. Proofs are in the appendix, which opens with a notation summary.

\subsection{Related Work}\label{sec:related}

\paragraph{Model selection under drift.}
\citet{han2024model} compare candidate models on adaptively chosen rolling windows, \citet{capponi2025nonstationarity} choose the model class and training window jointly, and \citet{mazzetto2023adaptive} choose a training window without knowing the size of the drift. All three choose which data to use. We keep every fold and weight them.

\paragraph{Weighting.}
\citet{barber2023conformal} use fixed weights in conformal prediction. For independent data, their coverage gap is at most twice a weighted sum of the total variation distances from each point to the test point. They note that under gradual drift, weights that decay geometrically with age will likely give a low coverage gap without knowledge of the exact rate of drift. In Theorem~\ref{thm:weighted_bound} the fold weights likewise multiply bounds on total variation distances, but our bound also has a noise term that grows as the weights concentrate.
\citet{jeong2025rider} and \citet{brock2026fastrates} weight past data in the training objective, and \citet{brock2026fastrates} balance learning error against drift. We weight fold scores used for selection. \citet{sugiyama2007iwcv} and \citet{brenning2026twcv} weight validation losses toward a target distribution that is known or sampled, and $P_{T+s}$ is neither. \citet{vamsikrishna2024ewkcv} propose exponentially weighted $k$-fold cross-validation for time series and evaluate it empirically.

\section{Setup}\label{sec:setup}

Data arrives in batches over periods $t = 1, \ldots, T$, and the model is deployed at a future time $T+s$, $s > 0$. The target is risk under the deployment distribution $P_{T+s}$, which is never observed.

Let $\mathcal{X}$, $\mathcal{Y}$, and $\mathcal{A}$ be measurable feature, response, and prediction spaces, and let $\ell : \mathcal{A} \times \mathcal{Y} \to \mathbb{R}$ be a measurable loss. For probability measures $P, Q$ on $\mathcal{X} \times \mathcal{Y}$, the total variation distance is
\[
    d_{\mathrm{TV}}(P,\, Q) \;=\; \sup_{A} |P(A) - Q(A)| \;\in\; [0, 1],
\]
the supremum over measurable sets. For any measurable $g$ with values in $[0, M]$, $|\E_P[g] - \E_Q[g]| \leq M\, d_{\mathrm{TV}}(P, Q)$.

\subsection{Distributional Drift}

\begin{description}
    \item[Assumption 1 (Latent Drift Path).] Let $(\Theta, d_\Theta)$ be a metric space of latent states, let 
    $\theta:\,[0,\infty)\to  \Theta$ be a continuous trajectory, and let $\{P_\vartheta\}_{\vartheta\in\Theta}$ be a family of probability measures on $\mathcal{X} \times \mathcal{Y}$. The deployment distribution is $P_{\theta(T+s)}$.

    \item[Assumption 2 (Independent Sampling).] At each integer time $t = 1, \ldots, T$, the data contains $B_t \geq 1$ samples $\{(X_{t,i},\, Y_{t,i})\}_{i=1}^{B_t} \overset{\mathrm{i.i.d.}}{\sim} P_{\theta(t)}$, drawn independently across $t$. All observations are therefore mutually independent, though not identically distributed.

    \item[Assumption 3 (H\"older Drift).] There exist constants $L > 0$ and $\alpha \in (0, 1]$ such that
    \[
        d_{\mathrm{TV}}(P_\vartheta,\, P_{\vartheta'}) \leq L\, d_\Theta(\vartheta,\, \vartheta')^{\alpha} \qquad \text{for all } \vartheta, \vartheta' \in \Theta.
    \]
\end{description}

\noindent Throughout the main text $\theta(t) = t$ on $\Theta = [0,\infty)$, so $P_{\theta(t)} = P_t$ and
\[
    d_{\mathrm{TV}}(P_t,\, P_{t'}) \leq L\, |t - t'|^{\alpha} \qquad \text{for all } t, t' \geq 0.
\]
A general $\theta$ covers recurring drift, such as a cycle $\theta(t) = (\cos(2\pi t/\tau), \sin(2\pi t/\tau))$, under which an old fold at the deployment phase can sit close to deployment (Appendix~\ref{app:general_theta}). Since $d_{\mathrm{TV}} \leq 1$, every drift bound below is truncated at $1$, written $a \wedge 1 = \min\{a, 1\}$.

\subsection{Hypothesis Classes}

Let $\Lambda$ be a set of model configurations. A configuration fixes everything chosen before seeing data, such as the model type and its structural hyperparameters; for example, a regression tree's depth, with the splits learned from data. For each $\lambda \in \Lambda$, let $\mathcal{F}_\lambda$ be the set of measurable functions $f : \mathcal{X} \to \mathcal{A}$ that $\lambda$ admits as its learned parameters vary.

\begin{description}
    \item[Assumption 4 (Bounded Loss).] There exists $M > 0$ such that $\ell(f(x), y) \in [0, M]$ for all $f \in \mathcal{F}_\lambda$, $\lambda \in \Lambda$, $x \in \mathcal{X}$, and $y \in \mathcal{Y}$.

    \item[Assumption 5 (Finite Configurations).] The set $\Lambda$ is finite, with $|\Lambda| = J$.
\end{description}

Grid and random search satisfy Assumption~5, and adaptive search such as Bayesian optimization does not (Section~\ref{sec:future}).

\subsection{Training Mixture}\label{sec:training_mixture}

The deployed model is fit on the most recent $h \geq 1$ periods. Pooling them gives the training mixture
\[
    Q_h \;=\; \frac{1}{N_h}\sum_{t=T-h+1}^{T} B_t\, P_t, \qquad N_h = \sum_{t=T-h+1}^{T} B_t,
\]
the distribution of a draw that picks period $t$ with probability proportional to $B_t$ and then draws from $P_t$. A fold that trains on periods $\{k, \ldots, k+h-1\}$ has training mixture
\[
    Q^{(k)} \;=\; \frac{1}{N_{\mathrm{train},k}}\sum_{j=k}^{k+h-1} B_j\, P_j.
\]
Each fold reserves $m$ periods for validation, so no fold's training window is that of $Q_h$.

\subsection{Deployment Risk and the Oracle}\label{sec:pop_min}

\begin{description}
    \item[Assumption 6 (Population Minimizer).] For each $\lambda \in \Lambda$, a minimizer
    \[
        f^\star_{\lambda, Q_h} \in \arg\min_{f \in \mathcal{F}_\lambda} \E_{Q_h}[\ell(f(X), Y)]
    \]
    exists.
\end{description}

\noindent This is the model $\lambda$ would produce from unlimited data from $Q_h$. Its deployment risk and the oracle configuration are
\[
    \mathcal{R}(\lambda) \;=\; \E_{P_{T+s}}[\ell(f^\star_{\lambda, Q_h}(X), Y)], \qquad \lambda^\star \in \arg\min_{\lambda \in \Lambda}\; \mathcal{R}(\lambda).
\]
Since $P_{T+s}$ is unknown, $\lambda^\star$ cannot be computed. The question is whether cross-validation on periods $1, \ldots, T$ selects a configuration $\hat\lambda$ whose deployment risk is close to the oracle's.

The oracle depends on $h$ through $Q_h$, so we hold $h$ fixed.

\begin{remark}[Selection error and estimation error]\label{rem:selection}
In practice one deploys $\hat{f}_{\hat{\lambda}}$, the fit of $\hat\lambda$ on the last $h$ periods. Its excess risk splits as
\[
    \E_{P_{T+s}}[\ell(\hat{f}_{\hat{\lambda}})] - \mathcal{R}(\lambda^\star) \;=\; \underbrace{\E_{P_{T+s}}[\ell(\hat{f}_{\hat{\lambda}})] - \mathcal{R}(\hat{\lambda})}_{\text{estimation error}} \;+\; \underbrace{\mathcal{R}(\hat{\lambda}) - \mathcal{R}(\lambda^\star)}_{\text{selection error}}.
\]
The estimation error arises however $\lambda$ is chosen. The theorems bound the selection error.
\end{remark}

\section{Sliding-Window Temporal Cross-Validation}\label{sec:cv}
\begin{figure}[H]
\centering
\begin{tikzpicture}[
    x=0.62cm, y=1cm,
    trainbar/.style={draw=blue!55!black, fill=blue!14, semithick},
    valbar/.style={draw=orange!75!black, fill=orange!30, semithick},
    lbl/.style={font=\scriptsize},
    tny/.style={font=\tiny},
]
\def\bh{0.30}
\def\dep{12}

\draw[dashed, black!45] (\dep,0.05) -- (\dep,4.20);
\node[lbl, black!60, above] at (\dep,4.17) {deployment};

\foreach \i/\y in {1/3.72, 2/3.20, 3/2.68, 4/2.16, 5/1.64, 6/1.12} {
    \filldraw[trainbar] ({\i-0.4},\y) rectangle ({\i+2+0.4},\y+\bh);
    \filldraw[valbar]   ({\i+3-0.4},\y) rectangle ({\i+4+0.4},\y+\bh);
    \node[lbl, anchor=east] at (0.35,\y+0.5*\bh) {Fold $\i$};
}
\node[tny] at (2.0,3.72+0.5*\bh) {train};
\node[tny] at (4.5,3.72+0.5*\bh) {val};

\draw[dotted, black!45] (-2.4,0.98) -- (11.0,0.98);

\filldraw[trainbar] (7.6,0.52) rectangle (10.4,0.52+\bh);
\node[tny] at (9.0,0.52+0.5*\bh) {train};
\node[lbl, anchor=east] at (0.35,0.52+0.5*\bh) {Final fit};

\draw[->, semithick, black!65] (-0.1,-0.05) -- (13.2,-0.05) node[right, font=\scriptsize] {time};
\foreach \t in {1,...,10} {
    \draw[black!45] (\t,0.02) -- (\t,-0.12);
    \node[tny, below=0pt] at (\t,-0.10) {\t};
}
\draw[black!45] (\dep,0.02) -- (\dep,-0.12);
\node[tny, below=0pt] at (\dep,-0.10) {$T{+}s$};
\end{tikzpicture}
\caption{Sliding-window cross-validation with $T = 10$, $h = 3$, and $m = 2$, so $K = 6$. Fold $k$ trains on periods $k, \ldots, k+h-1$ (blue) and validates on the next $m$ (orange). The selected $\hat\lambda$ is then refit on the last $h$ periods, the window of $Q_h$, and deployed at $T+s$ (dashed).}
\label{fig:folds}
\end{figure}
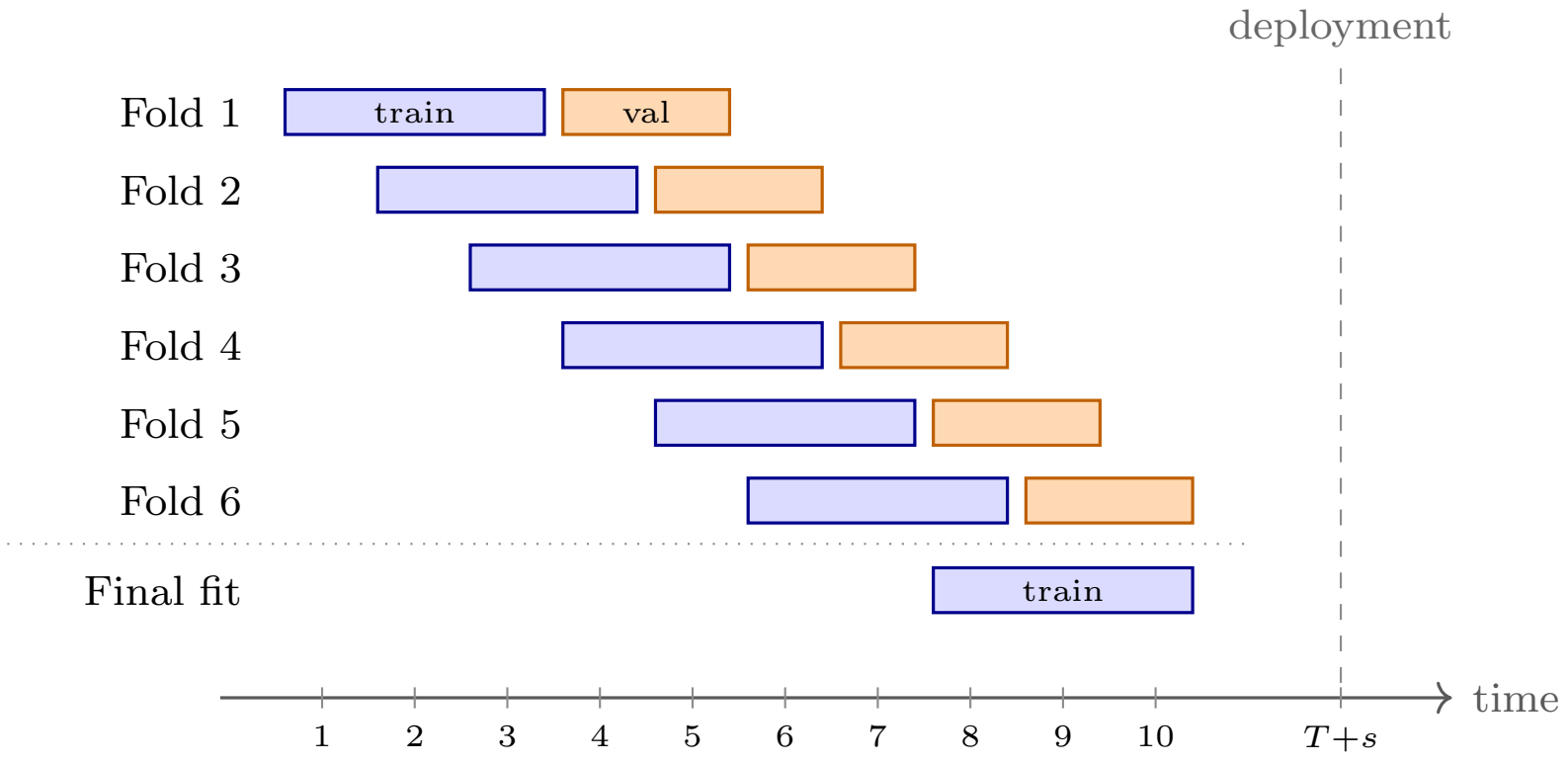

Each fold repeats the deployment fit on an earlier stretch of the record. It trains on $h$ consecutive periods, the same width as $Q_h$, and validates on the $m \geq 1$ periods that follow, with $h + m \leq T$. Rolling this window across the record gives $K = T - h - m + 1$ folds. At fold $k = 1, \ldots, K$, the training and validation sets are
\begin{align*}
    \mathcal{D}_{\mathrm{train},k} &= \{(X_{t,i},\, Y_{t,i}) : k \leq t \leq k + h - 1,\; 1 \leq i \leq B_t\}, \\
    \mathcal{D}_{\mathrm{val},k} &= \{(X_{t,i},\, Y_{t,i}) : k + h \leq t \leq k + h + m - 1,\; 1 \leq i \leq B_t\},
\end{align*}
with sizes $N_{\mathrm{train},k} = \sum_{t=k}^{k+h-1} B_t$ and $N_{\mathrm{val},k} = \sum_{t=k+h}^{k+h+m-1} B_t$. For each $\lambda \in \Lambda$ and each fold $k$, fit an empirical risk minimizer, which we assume exists, and score it on the validation block:
\begin{align*}
    \hat{f}_{\lambda,k} &\in \arg\min_{f \in \mathcal{F}_\lambda} \frac{1}{N_{\mathrm{train},k}} \sum_{t=k}^{k+h-1} \sum_{i=1}^{B_t} \ell(f(X_{t,i}),\, Y_{t,i}), \\
    \widehat{\mathrm{CV}}_k(\lambda) &= \frac{1}{N_{\mathrm{val},k}} \sum_{t=k+h}^{k+h+m-1} \sum_{i=1}^{B_t} \ell(\hat{f}_{\lambda,k}(X_{t,i}),\, Y_{t,i}).
\end{align*}
Standard sliding-window cross-validation averages the folds equally, and the weighted version uses weights $w_k \geq 0$ with $\sum_k w_k = 1$, fixed before the data is seen:
\[
    \widehat{\mathrm{CV}}(\lambda) = \frac{1}{K} \sum_{k=1}^{K} \widehat{\mathrm{CV}}_k(\lambda),
    \qquad
    \widehat{\mathrm{CV}}_w(\lambda) = \sum_{k=1}^{K} w_k\, \widehat{\mathrm{CV}}_k(\lambda).
\]
Each selects the minimizer over $\Lambda$, written $\hat{\lambda}$ in both cases, which is refit on the window of $Q_h$ for deployment. Figure~\ref{fig:folds} illustrates the procedure.

\section{Main Results}\label{sec:results}

Theorem~\ref{thm:upper_bound} bounds the excess deployment risk of equal-weight cross-validation, and Theorem~\ref{thm:weighted_bound} extends the bound to any fold weights. Both need one assumption beyond Assumptions~1--6. Cross-validation scores fitted models, not the best model in each class, so the bounds need the fit to approach that best model as its training block grows. Assumption~7 gives this by bounding how much better any model in $\mathcal{F}_\lambda$ can look on a sample than under the mixture the sample was drawn from.

\begin{description}
    \item[Assumption 7 (Statistical Complexity).] For each $\lambda\in \Lambda$, there exists $C_\lambda > 0$ with the following property. Let $(X_1,Y_1),\ldots,(X_N,Y_N)$ be independent observations, the $i$th drawn from $P_{t_i}$ for some period $t_i$, and let $Q = \sum_j \frac{B_j}{N} P_j$, where $B_j$ counts the observations from period $j$. Then a minimizer $f^\star_{\lambda, Q} \in \arg\min_{f \in \mathcal{F}_\lambda} \E_Q[\ell(f)]$ exists, and with $\hat{\E}[\ell(f)] = \frac{1}{N}\sum_{i=1}^{N} \ell(f(X_i),\, Y_i)$ the empirical risk,
    \[
    \E\Bigl[\sup_{f \in \mathcal{F}_\lambda} \bigl(\E_Q[\ell(f)] - \hat{\E}[\ell(f)]\bigr)\Bigr] \leq \frac{C_\lambda}{\sqrt{N}}.
    \]
    We take the supremum to be measurable.
\end{description}

By the symmetrization step in the proof of Theorem~3.3 of \citet{mohri2018foundations}, which carries over from i.i.d.\ to independent data, the left side of Assumption~7 is at most twice the Rademacher complexity of the loss class, so the assumption holds whenever that complexity is at most $C_\lambda/(2\sqrt{N})$, as it is for a finite class or for bounded linear predictors with a Lipschitz loss. The following lemma turns Assumption~7 into a bound on how far the fit falls short of the best model in $\mathcal{F}_\lambda$.

\begin{lemma}[Estimation error]\label{lem:erm}
Let Assumptions~4 and~7 hold, fix $\lambda \in \Lambda$, and let the observations $(X_i,Y_i)$ and the mixture $Q$ be as in Assumption~7. Let $\hat{f}$ minimize the empirical risk $\hat{\E}[\ell(f)]$ over $\mathcal{F}_\lambda$. For every $\delta \in (0,1)$, with probability at least $1-\delta$,
\[
\E_Q\bigl[\ell(\hat{f})\bigr] - \E_Q\bigl[\ell(f^\star_{\lambda,Q})\bigr] \leq \frac{C_\lambda}{\sqrt{N}} + M\sqrt{\frac{2\log(2/\delta)}{N}},
\]
where $\E_Q[\ell(\hat{f})]$ is the risk of the fitted function under $Q$, with $\hat f$ held fixed.
\end{lemma}

The proof, in Appendix~\ref{app:erm}, follows that of Theorem~1 of \citet{mohri2012domain}, with the training mixture as the target.

\subsection{One Fold}

Every drift term in the bounds is the distance from one period to a mixture of periods. The following lemma bounds it.

\begin{lemma}[Drift to a mixture]\label{lem:drift}
Let Assumptions 1--3 hold, and let $Q = \sum_j \nu_j P_j$ be a mixture of periods with weights $\nu_j \geq 0$ summing to one. For any time $t \geq 0$,
\[
    d_{\mathrm{TV}}(P_t,\, Q) \;\leq\; \sum_j \nu_j \bigl(L|t-j|^\alpha \wedge 1\bigr).
\]
\end{lemma}

\begin{proof}
By joint convexity of total variation distance, $d_{\mathrm{TV}}(P_t, Q) \leq \sum_j \nu_j\, d_{\mathrm{TV}}(P_t, P_j)$. Assumption~3 bounds the $j$th distance by $L|t-j|^\alpha$, and each distance is at most $1$.
\end{proof}

For a fixed configuration, the next lemma bounds the gap between deployment risk and the mean score of one fold in each direction. Both theorems combine it over the folds.

\begin{lemma}[One fold]\label{lem:fold}
Let Assumptions 1--7 hold, and fix $\lambda \in \Lambda$ and a fold $k$ with validation periods $V_k = \{k+h, \ldots, k+h+m-1\}$. Write $\pi_t = B_t/N_{\mathrm{val},k}$ for the share of the validation block from period $t$, and $\omega_j = B_j/N_{\mathrm{train},k}$ and $\gamma_{j'} = B_{j'}/N_h$ for the weights of periods $j$ and $j'$ in the mixtures $Q^{(k)}$ and $Q_h$. With equal batch sizes, $\pi_t = 1/m$ and $\omega_j = \gamma_{j'} = 1/h$. Set
\begin{align*}
q &= \sum_{j'=T-h+1}^{T}\gamma_{j'}\bigl(L(T+s-j')^\alpha \wedge 1\bigr),
&
u_k &= \sum_{t\in V_k}\pi_t\sum_{j'=T-h+1}^{T}\gamma_{j'}\bigl(L|t-j'|^\alpha \wedge 1\bigr),\\
r_k &= \sum_{j=k}^{k+h-1}\omega_j\bigl(L(T+s-j)^\alpha \wedge 1\bigr),
&
b_k &= \sum_{t\in V_k}\pi_t\sum_{j=k}^{k+h-1}\omega_j\bigl(L|t-j|^\alpha \wedge 1\bigr).
\end{align*}
Write
\[
\mathrm{CV}_k(\lambda) = \sum_{t\in V_k}\pi_t\,\E_{P_t}[\ell(\hat{f}_{\lambda,k})]
\]
for the mean of the fold score $\widehat{\mathrm{CV}}_k(\lambda)$ given the training block $\mathcal{D}_{\mathrm{train},k}$.

\medskip
\noindent\textup{\textbf{Underestimate.}} For every draw of $\mathcal{D}_{\mathrm{train},k}$,
\[
\mathcal{R}(\lambda)-\mathrm{CV}_k(\lambda)
\leq \min\bigl(M,\; Mq + Mu_k\bigr).
\]

\medskip
\noindent\textup{\textbf{Overestimate.}} For every $\delta \in (0,1)$, with probability at least $1-\delta$ over $\mathcal{D}_{\mathrm{train},k}$,
\[
\mathrm{CV}_k(\lambda)-\mathcal{R}(\lambda)
\leq \min\Biggl(M,\; Mb_k + Mr_k + \frac{C_\lambda}{\sqrt{N_{\mathrm{train},k}}} + M\sqrt{\frac{2\log(2/\delta)}{N_{\mathrm{train},k}}}\Biggr).
\]
\end{lemma}

Deployment risk $\mathcal{R}(\lambda)$ scores the population model $f^\star_{\lambda,Q_h}$ under $P_{T+s}$, and the mean fold score $\mathrm{CV}_k(\lambda)$ scores the fit $\hat f_{\lambda,k}$ under the validation periods. To compare them, the proof moves both to a distribution under which we can bound how much worse one model is than the other, and by Lemma~\ref{lem:drift} each move costs a drift term. For the underestimate, that distribution is $Q_h$. The population model is best in $\mathcal{F}_\lambda$ under $Q_h$, so it is no worse than the fit, and the moves cost $q$, the drift from deployment to $Q_h$, and $u_k$, the drift from $Q_h$ to the validation periods. For the overestimate, it is the fold's training mixture $Q^{(k)}$. Under $Q^{(k)}$ the fit is worse than the best model in $\mathcal{F}_\lambda$, and so worse than the population model, by at most its estimation error (Lemma~\ref{lem:erm}), and the moves cost $b_k$, the drift from the validation periods to $Q^{(k)}$, and $r_k$, the drift from $Q^{(k)}$ to deployment. The proof is in Appendix~\ref{app:fold}.

The theorems apply the first bound at the selected $\hat\lambda$ and the second at the fixed oracle $\lambda^\star$. Since $\hat\lambda$ depends on the data, it matters that the first bound holds for every draw.

\subsection{Equal Weights}

\begin{theorem}[Upper Bound on Excess Deployment Risk]\label{thm:upper_bound}
Let Assumptions 1--7 hold. Let $\hat{\lambda}$ be the configuration selected by equal-weight sliding-window cross-validation (Section~\ref{sec:cv}) with training width $h$, validation width $m$, and $K = T - h - m + 1$ folds. Let $q$, $u_k$, $b_k$, and $r_k$ be the drift terms in Lemma~\ref{lem:fold}, and write $C_{\max} = \max_{\lambda \in \Lambda} C_\lambda$ and $N_{\mathrm{val},\min} = \min_k N_{\mathrm{val},k}$. Then for any $\delta \in (0, 1)$, with probability at least $1 - \delta$,
\begin{equation}\label{eq:thm1}
\boxed{\;
\begin{aligned}
\mathcal{R}(\hat{\lambda}) - \mathcal{R}(\lambda^\star)
&\;\leq\;
\frac{1}{K}\sum_{k=1}^{K}\min\bigl(M,\; Mq + Mu_k\bigr) \\
&\quad + \frac{1}{K}\sum_{k=1}^{K}\min\Biggl(M,\; Mb_k + Mr_k + \frac{C_{\max}}{\sqrt{N_{\mathrm{train},k}}} + M\sqrt{\frac{2\log(6K/\delta)}{N_{\mathrm{train},k}}}\Biggr) \\
&\quad + M\sqrt{\frac{m\log(3J/\delta)}{2KN_{\mathrm{val},\min}}} + M\sqrt{\frac{m\log(3/\delta)}{2KN_{\mathrm{val},\min}}}.
\end{aligned}
\;}
\end{equation}
\end{theorem}

\begin{proof}[Proof sketch]
The proof is in Appendix~\ref{app:thm1}. Since $\hat\lambda$ minimizes the score, the excess risk is at most how far the score underestimates the risk of $\hat\lambda$ plus how far it overestimates the risk of $\lambda^\star$. Lemma~\ref{lem:fold} bounds each fold's gap in both directions. The folds overlap, so for the validation noise we sum the deviations period by period. Each period's term has mean zero given the earlier periods, since every fold that validates on it was fit on earlier data, so Azuma's inequality applies without independence between folds.
\end{proof}

The two sums are the drift and estimation terms of Lemma~\ref{lem:fold}, averaged over folds, and the last two terms are validation noise. Only the third pays the union bound over the $J$ configurations, since the selected configuration depends on the data.

Equal weights give the oldest folds, which sit furthest from deployment, as much weight as the newest. Every fold whose validation block ends at least $L^{-1/\alpha}$ periods before the window of $Q_h$ begins has $u_k = r_k = 1$, so its term in each sum is the cap $M$. On a long record almost every fold is of this kind, so the drift terms saturate and the bound is no better than $M$, which the excess risk never exceeds anyway. Theorem~\ref{thm:weighted_bound} allows weights that shrink the share of these folds, including a hard cutoff that drops them.

\subsection{Arbitrary Weights}\label{sec:weighted_bound}

\begin{theorem}[Weighted Upper Bound on Excess Deployment Risk]\label{thm:weighted_bound}
Let Assumptions 1--7 hold. Let $\hat{\lambda}$ be the configuration selected by weighted sliding-window cross-validation (Section~\ref{sec:cv}) with training width $h$, validation width $m$, $K = T - h - m + 1$ folds, and weights $w_k \geq 0$ with $\sum_{k=1}^K w_k = 1$ that do not depend on the data. Let $q$, $u_k$, $b_k$, and $r_k$ be the drift terms in Lemma~\ref{lem:fold}, write $C_{\max} = \max_{\lambda \in \Lambda} C_\lambda$, $N_{\mathrm{val},\min} = \min_k N_{\mathrm{val},k}$, and $\|w\|_2 = (\sum_{k=1}^K w_k^2)^{1/2}$. Then for any $\delta \in (0, 1)$, with probability at least $1 - \delta$,
\begin{equation}\label{eq:thm2}
\boxed{\;
\begin{aligned}
\mathcal{R}(\hat{\lambda}) - \mathcal{R}(\lambda^\star)
&\;\leq\;
\sum_{k=1}^{K} w_k\min\bigl(M,\; Mq + Mu_k\bigr) \\
&\quad + \sum_{k=1}^{K} w_k\min\Biggl(M,\; Mb_k + Mr_k + \frac{C_{\max}}{\sqrt{N_{\mathrm{train},k}}} + M\sqrt{\frac{2\log(6K/\delta)}{N_{\mathrm{train},k}}}\Biggr) \\
&\quad + M\|w\|_2\sqrt{\frac{m\log(3J/\delta)}{2N_{\mathrm{val},\min}}} + M\|w\|_2\sqrt{\frac{m\log(3/\delta)}{2N_{\mathrm{val},\min}}}.
\end{aligned}
\;}
\end{equation}
\end{theorem}

\begin{proof}[Proof sketch]
The proof is in Appendix~\ref{app:thm2}. It is the proof of Theorem~\ref{thm:upper_bound} with the fold average replaced by the weights $w_k$. Each period is in the validation block of at most $m$ folds, which gives the factor $\sqrt{m}\,\|w\|_2$.
\end{proof}

The weights multiply the capped fold terms, so weight on recent folds, whose $u_k$ and $r_k$ are small, shrinks the drift sums. The noise terms grow with $\|w\|_2$, which is smallest for equal weights, $\|w\|_2 = 1/\sqrt{K}$, where the bound recovers Theorem~\ref{thm:upper_bound}, and largest, $\|w\|_2 = 1$, for all weight on one fold. The fold sums are linear in $w$ and the noise terms are convex, so minimizing the bound over the weights is a convex problem (Section~\ref{sec:future}).

The balance depends on the number of observations per period. The noise terms shrink as $N_{\mathrm{val},\min}$ grows and the drift sums do not, so with many observations per period the bound favors putting nearly all weight on the folds with the smallest drift terms, which are among the most recent. With few observations per period, spreading the weight over more folds lowers the bound. Section~\ref{sec:apps} shows both sides, on earnings and flight delays.

\section{Simulation}\label{sec:sim}

We simulate a record in which the predictive feature changes slowly and ask which configuration cross-validation selects as older history is added. Theorems~\ref{thm:upper_bound} and~\ref{thm:weighted_bound} suggest that equal weights drift toward the configuration that suited the distant past, decaying weights should not, and weights on too few folds should pay in noise.

Theorem~\ref{thm:weighted_bound} has two extremes. Its noise term is smallest under equal weights, and its drift terms are smallest when the weight sits on the folds nearest deployment. We study weights that move between these extremes with one parameter, geometric in fold age with ratio $\rho \geq 0$, $w_k = \rho^{K-k}/\sum_{l=1}^{K}\rho^{K-l}$, with $0^0 = 1$. Equal weights are $\rho = 1$, and $\rho = 0$ puts all weight on the last fold, which is the usual holdout on the most recent period. Between them, the half-life rule $\rho = 2^{-1/h}$ halves a fold's weight every $h$ periods and needs no knowledge of the drift. Values $\rho > 1$ give weights that grow with age.

Define the features as $X \sim U[-1,1]^4$, and in period $t$
\[
Y = \sum_{j=1}^{4} \beta_j(a)\, X_j + \varepsilon, \qquad \varepsilon \sim U[-1,1], \qquad \beta_j(a) = e^{-(j - c(a))^2/2}, \qquad c(a) = \max(1,\, 4 - a/20),
\]
where $a = T+1-t$ is the age of period $t$. Configuration $j$ is least squares on $X_j$ with an intercept, predictions clipped to $[-3,3]$. Each period has $B_t = 10$ observations, $h = 5$, $m = 1$, and deployment is at $T+1$. We report $\mathcal{R}(\hat\lambda) - \mathcal{R}(\lambda^\star)$, with $\mathcal{R}$ from large-sample fits, averaged over 1000 replications.

The relevant feature is the $X_j$ with $j$ closest to $c(a)$. It is $X_4$ in the last 10 periods, $X_3$ from 10 to 30 periods back, $X_2$ from 30 to 50, and $X_1$ before that. The drift is Lipschitz in total variation ($\alpha = 1$), and a longer record only adds older history. Clipping bounds the loss (Assumption~4).

\begin{figure}[H]
\centering
\includegraphics[width=\textwidth]{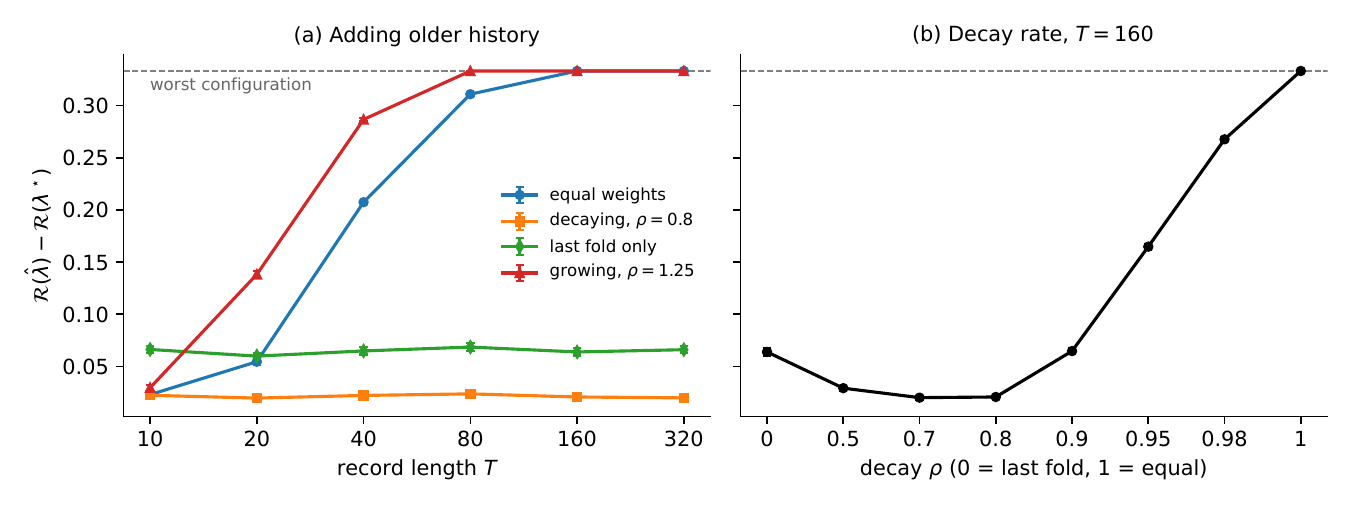}
\caption{Excess deployment risk of the selected configuration. (a) As older history is added, equal weights and weights that grow with age drift to the worst configuration; decaying weights and the last fold alone do not. (b) At $T = 160$, concentrated weights pay in noise and spread weights in drift. Error bars are $\pm 1$ standard error.}
\label{fig:sim}
\end{figure}

In Figure~\ref{fig:sim}(a), equal weights rise to the worst configuration, $X_1$, as old folds come to dominate (Theorem~\ref{thm:upper_bound}), and growing weights, $\rho = 1.25$, get there sooner. Decaying weights, $\rho = 0.8$, stay flat. The last fold alone is flat but three times higher, and at $T = 10$ worse than equal weights, $0.066$ against $0.023$.

In Figure~\ref{fig:sim}(b), at $T = 160$, every $\rho < 1$ beats equal weights, and the excess risk is U-shaped in $\rho$, smallest at $\rho = 0.7$ to $0.8$ with a third of the last-fold value. The last fold sometimes picks $X_1$ or $X_2$ by chance, while $\rho = 0.9$ picks $X_3$ a third of the time, from folds where $X_3$ was best. This is the trade-off of Theorem~\ref{thm:weighted_bound}. Choosing $\rho$ without knowing the drift is open (Section~\ref{sec:future}).

\section{Applications}\label{sec:apps}

We compare equal weights, the last fold alone, and the half-life rule (Section~\ref{sec:sim}) on two real records in which the best configuration changes. The first two fail in opposite directions. When a change in the best configuration persists, as on U.S. earnings, the last fold follows it and equal weights lag behind. When changes do not last, as on U.S. flight delays, the last fold chases them and equal weights average them out. The half-life rule, fixed in advance, is never the worst of the three in either record.

In each application we treat all the data of a period as its population $P_t$. This lets us compute $\mathcal{R}$ without sampling error, since we can fit each configuration on all the data of the window, with each period weighted equally as in $Q_h$, and score it on all the data of the deployment period. Cross-validation sees only $B_t$ observations per period, drawn from $P_t$ at random with replacement, so they are independent (Assumption~2). For simplicity and computation, each configuration is a least-squares regression on a single characteristic. We use $h = 5$ and $m = 1$, deploy at $T+1$ in every period after the first $h+1$, use the weights of Section~\ref{sec:sim}, and report $\mathcal{R}(\hat\lambda) - \mathcal{R}(\lambda^\star)$ over 200 replications. The constants $L$ and $\alpha$ are unknown, so we cannot evaluate the bound. Even its noise terms alone are about 200 times the excess risk we observe, since they come from worst-case inequalities. We use the bound for direction, to see which weights it favors as $B_t$ and the drift vary. Appendices~\ref{app:cps} and~\ref{app:flights} give the samples, the protocols, and further results. Code for the simulation and both applications is available as ancillary files with this paper.

\subsection{Earnings}\label{sec:cps}

\begin{figure}[t]
\centering
\includegraphics[width=\textwidth]{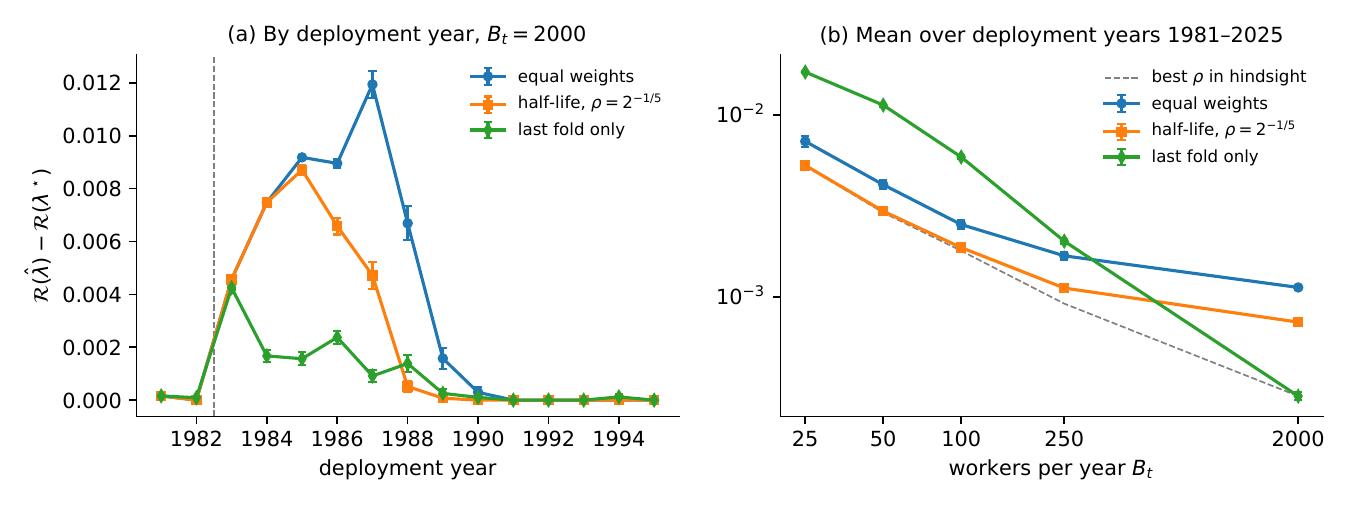}
\caption{Excess deployment risk on CPS earnings. (a) By deployment year, 1981 to 1995, with 2000 workers per year; the dashed line separates 1982, the last year sex is best, from 1983, the first year education is. (b) Mean over deployment years. The dashed line is the best $\rho$ in hindsight. Error bars are $\pm 1$ standard error over 200 replications.}
\label{fig:cps}
\end{figure}

The periods are the earnings years 1975 to 2025 of the Current Population Survey \citep{ipumscps2026}, $Y$ is the log hourly wage of full-time workers, and the characteristics are education, sex, age, race and Hispanic origin, and marital status. Under $\mathcal{R}$, sex is the best configuration through 1982 and education from 1983 on, so after 1983 the old folds point the wrong way. With 2000 workers per year, one year of validation data is enough to rank the configurations, and since the change persists, the newest fold is the most informative. The last fold is the first to follow the change. It pays mostly when deploying in 1983, since every fold then validates on a year up to 1982, when sex was still best (Figure~\ref{fig:cps}(a)). Any weight on older folds keeps folds that still favor sex in the score, so equal weights take until 1990 to follow the change and the half-life rule until 1988. With 100 or fewer workers per year, one year is no longer enough. The last fold's single validation year is noisy, and it does three to four times worse than the half-life rule (Figure~\ref{fig:cps}(b)). The weights that minimize the bound move the same way, toward recent folds as $B_t$ grows, since its noise terms shrink with $B_t$ and its drift sums do not.

\subsection{Flight Delays}\label{sec:flights}

\begin{figure}[t]
\centering
\includegraphics[width=\textwidth]{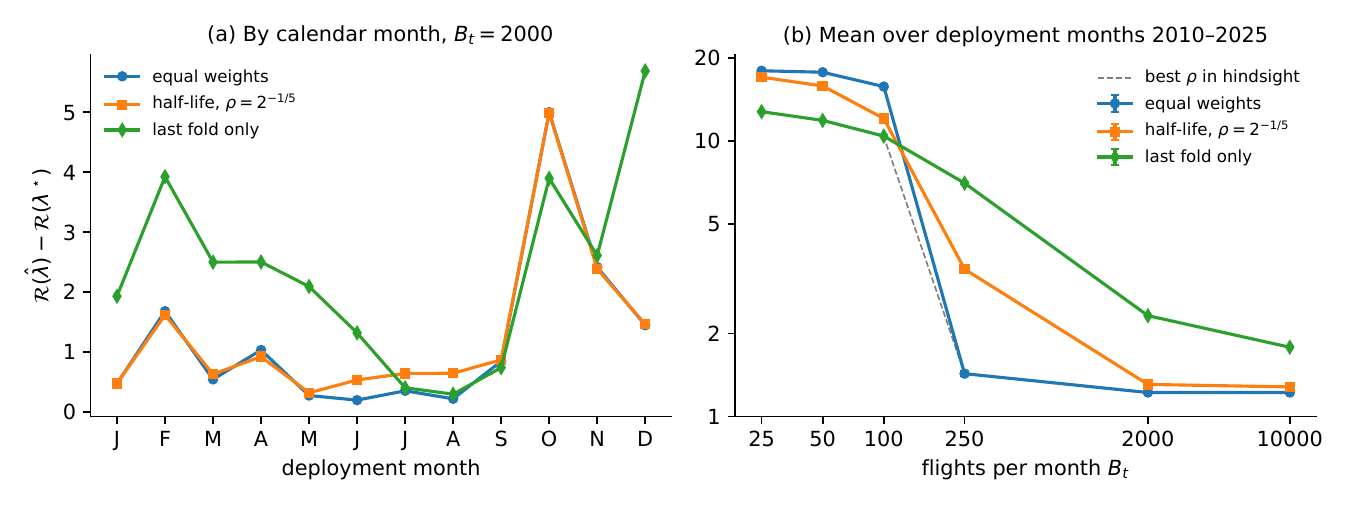}
\caption{Excess deployment risk on flight delays. (a) Mean by calendar month of deployment, with 2000 flights per month. (b) Mean over deployment months. The dashed line is the best $\rho$ in hindsight. Error bars are $\pm 1$ standard error over 200 replications.}
\label{fig:flights}
\end{figure}

The periods are the months January 2010 to December 2025 of the on-time performance data \citep{bts2026ontime}, $Y$ is the arrival delay in minutes, and the characteristics are departure hour, carrier, origin airport, destination airport, and day of the week. Under $\mathcal{R}$, departure hour is best in 137 of the 186 deployment months. In the others a different configuration is best by a small margin, usually for a month or two. So the newest month is a poor guide to the next, and the best fixed choice is departure hour. With 2000 flights per month, equal weights average over many months and select departure hour in every deployment month. The last fold switches to whichever configuration led the latest month, and it loses to equal weights in every calendar month from December to June (Figure~\ref{fig:flights}(a)). It gains only in September and October, when the summer lead of departure hour ends and the newest fold is the first to see it.

From 250 flights per month up, equal weights are best and the last fold is worst (Figure~\ref{fig:flights}(b)). The last fold loses even when every fold is scored on all flights of its months (Appendix~\ref{app:flights}), so its loss comes from the newest month pointing the wrong way, not from noise. At 100 or fewer flights per month every weighting does far worse, since each fold trains on too few flights to fit the configurations with many groups (Appendix~\ref{app:flights}).

The half-life rule is within $0.0004$ of the best $\rho$ in hindsight on earnings and within $0.09$ on flights from 2000 flights per month up, but it is $2.4$ times the best at 250. In terms of Theorem~\ref{thm:weighted_bound}, drift on earnings grows with the gap between periods, which the drift terms penalize, while drift on flights behaves like $\alpha$ near 0, so the drift terms give no reason to favor recent folds.

\section{Future Work}\label{sec:future}

\paragraph{Choosing the weights.}
Minimizing the right-hand side of Theorem~\ref{thm:weighted_bound} over the simplex is a convex problem, but it minimizes a bound, not the risk, and depends on $L$ and $\alpha$, which are unknown and would have to be adapted to, as \citet{han2024model} do for the window. The applications show that the right $\rho$ depends on the record. In hindsight it is near 0 on earnings with 2000 workers per year and near 1 on flight delays with 250 or more flights per month. Choosing $\rho$ from the data, for example by how each value would have done at earlier deployments, needs the noise terms to hold for every candidate, which a union bound over a finite grid gives at the cost of the logarithm of the grid size.

\paragraph{Regime rather than age.}
Decaying weights assume that distributions drift further apart with time, as under $\theta(t) = t$. Under a recurring latent path this fails, and good weights require knowing or estimating the path $\theta$, not only $L$ and $\alpha$. There, an old fold can sit close to deployment, which shrinks its $r_k$ (Appendix~\ref{app:general_theta}). Its $u_k$ still compares its validation periods with the recent window of $Q_h$, so an old fold helps only when the recent window shares its regime. On flight delays, weighting folds by calendar month did no better than equal weights, since the short leads there rarely recur in the same month a year later (Appendix~\ref{app:flights}).

\paragraph{Continuous and adaptive tuning.}
Assumption~5 enters only through the union bound over configurations, as $\log J$ in the noise terms. For a continuous set of configurations, such as a regularization strength in an interval, a covering argument would replace $\log J$ by the logarithm of a covering number, provided the fold scores vary smoothly with $\lambda$. Adaptive search picks each configuration from earlier scores, so the candidates depend on the validation data, as tuned weights do. Covering it needs a bound that holds uniformly over the whole space, or held-out data for the final comparison.

\paragraph{Dependent observations.}
Assumption~2 rules out dependence, which fails for panels such as repeated observations of the same borrowers. Under a nonstationary version of the physical dependence measure of \citet{wu2005nonlinear}, the drift terms carry over unchanged, but the concentration steps and the conditioning on each training block must be rebuilt.

\paragraph{Evaluation.}
The simulation of Section~\ref{sec:sim} and the applications of Section~\ref{sec:apps} show the trade-off but do not test the bounds numerically. In the applications the small samples per period come from subsampling. Data with few observations per period by nature, such as weekly case counts in epidemic forecasting, would test the noise side of the trade-off directly.

\bibliographystyle{plainnat}
\bibliography{references}

\clearpage
\appendix

\section{Notation Summary}\label{app:notation}

\begin{table}[H]
\centering
\caption{Notation: spaces, data, and model selection.}
\label{tab:notation}
\renewcommand{\arraystretch}{1.15}
\begin{tabular}{@{}cl@{}}
\toprule
\textbf{Symbol} & \textbf{Description} \\
\midrule
\multicolumn{2}{@{}l}{\emph{Spaces and loss}} \\[2pt]
$\mathcal{X},\; \mathcal{Y},\; \mathcal{A}$ & Feature, response, and prediction spaces \\
$\ell \colon \mathcal{A} \times \mathcal{Y} \to \mathbb{R}$ & Loss function \\
$M$ & Uniform upper bound on the loss (Assumption~4) \\[4pt]
\multicolumn{2}{@{}l}{\emph{Distributional drift}} \\[2pt]
$P_t$ & Data-generating distribution at time $t$ \\
$\Theta,\; \theta(t)$ & Latent state space and trajectory \\
$L,\; \alpha$ & H\"older constant and exponent (Assumption~3) \\
$d_{\mathrm{TV}}$ & Total variation distance \\[4pt]
\multicolumn{2}{@{}l}{\emph{Observed data}} \\[2pt]
$T$ & Number of observed time periods \\
$B_t$ & Number of observations (batch size) at time $t$ \\
$s$ & Deployment horizon; deployment distribution is $P_{T+s}$ \\[4pt]
\multicolumn{2}{@{}l}{\emph{Model selection}} \\[2pt]
$\Lambda,\; J = |\Lambda|$ & Set of model configurations and its cardinality \\
$\lambda,\; \hat{\lambda},\; \lambda^\star$ & A configuration, CV-selected, and oracle \\
$\mathcal{F}_\lambda$ & Hypothesis class for configuration $\lambda$ \\
$f^\star_{\lambda, Q}$ & Population minimizer of $\mathcal{F}_\lambda$ under distribution $Q$ \\
$\hat{f}_{\lambda,k}$ & ERM estimator trained on fold-$k$ data \\
$C_\lambda,\; C_{\max}$ & Complexity constant of Assumption~7, and $\max_\lambda C_\lambda$ \\
\bottomrule
\end{tabular}
\end{table}

\begin{table}[H]
\centering
\caption{Notation: cross-validation and risk.}
\label{tab:notation2}
\renewcommand{\arraystretch}{1.15}
\begin{tabular}{@{}cl@{}}
\toprule
\textbf{Symbol} & \textbf{Description} \\
\midrule
\multicolumn{2}{@{}l}{\emph{Cross-validation}} \\[2pt]
$h,\; m$ & Training and validation window widths \\
$K$ & Number of folds, $K = T - h - m + 1$ \\
$Q_h$ & Training mixture over the most recent $h$ periods \\
$Q^{(k)}$ & Training mixture over the $h$ periods starting at $k$ \\
$N_h,\; N_{\mathrm{train},k},\; N_{\mathrm{val},k}$ & Sample sizes: full window, fold training, fold validation \\
$N_{\mathrm{val},\min}$ & $\min_{1 \leq k \leq K} N_{\mathrm{val},k}$ \\
$\widehat{\mathrm{CV}}_k(\lambda)$ & Fold-$k$ CV estimate \\
$\widehat{\mathrm{CV}}(\lambda)$ & Averaged CV estimate over all folds \\
$\mathrm{CV}_k(\lambda),\; \mathrm{CV}(\lambda)$ & Means of $\widehat{\mathrm{CV}}_k(\lambda)$ and $\widehat{\mathrm{CV}}(\lambda)$ given the training blocks (Lemma~\ref{lem:fold}) \\
$V_k$ & Validation periods of fold $k$, $\{k+h, \ldots, k+h+m-1\}$ \\
$\pi_t,\; \omega_j,\; \gamma_{j'}$ & Period weights in validation, in $Q^{(k)}$, and in $Q_h$ (Lemma~\ref{lem:fold}) \\
$w_k$ & Weight on fold $k$; $w_k \geq 0$, $\sum_k w_k = 1$ \\
$\widehat{\mathrm{CV}}_w(\lambda)$ & Weighted CV estimate, $\sum_k w_k \widehat{\mathrm{CV}}_k(\lambda)$ \\[4pt]
\multicolumn{2}{@{}l}{\emph{Risk}} \\[2pt]
$\mathcal{R}(\lambda)$ & Deployment risk, $\E_{P_{T+s}}[\ell(f^\star_{\lambda, Q_h})]$ \\
$q,\; u_k,\; b_k,\; r_k$ & Drift terms in Lemma~\ref{lem:fold} \\
$\|w\|_2$ & $(\sum_k w_k^2)^{1/2}$, the factor in the noise terms of Theorem~\ref{thm:weighted_bound} \\
\bottomrule
\end{tabular}
\end{table}
\newpage

\section{Proof of Lemma~\ref{lem:erm}}\label{app:erm}

\begin{proof}[Proof of Lemma~\ref{lem:erm}]
Write $f^\star = f^\star_{\lambda,Q}$. Adding and subtracting the empirical risks of $\hat f$ and $f^\star$,
\[
\E_Q[\ell(\hat f)] - \E_Q[\ell(f^\star)]
= \bigl(\E_Q[\ell(\hat f)] - \hat{\E}[\ell(\hat f)]\bigr)
+ \bigl(\hat{\E}[\ell(\hat f)] - \hat{\E}[\ell(f^\star)]\bigr)
+ \bigl(\hat{\E}[\ell(f^\star)] - \E_Q[\ell(f^\star)]\bigr).
\]
The middle term is at most zero, since $\hat f$ minimizes the empirical risk and $f^\star \in \mathcal{F}_\lambda$. We bound the other two, each with failure probability $\delta/2$, by McDiarmid's and Hoeffding's inequalities \citep[Theorems~D.8 and~D.2]{mohri2018foundations}, which need only independence.

\emph{First term.} The fit depends on the sample, so we bound the first term by
\[
\Phi = \sup_{f \in \mathcal{F}_\lambda} \bigl(\E_Q[\ell(f)] - \hat{\E}[\ell(f)]\bigr),
\]
whose mean is at most $C_\lambda/\sqrt{N}$ by Assumption~7. Replacing one observation changes one loss in $[0,M]$ with weight $1/N$, so it moves each term inside the supremum, and hence $\Phi$, by at most $c_i = M/N$. By McDiarmid's inequality, for $\varepsilon > 0$,
\[
P\bigl(\Phi - \E[\Phi] \geq \varepsilon\bigr) \leq \exp\Biggl(-\frac{2\varepsilon^2}{\sum_{i=1}^{N} c_i^2}\Biggr) = \exp\Biggl(-\frac{2\varepsilon^2}{N(M/N)^2}\Biggr) = \exp\Biggl(-\frac{2N\varepsilon^2}{M^2}\Biggr).
\]
Setting the right side to $\delta/2$ and solving gives $\varepsilon = M\sqrt{\log(2/\delta)/(2N)}$. So with probability at least $1-\delta/2$,
\[
\Phi \leq \E[\Phi] + M\sqrt{\frac{\log(2/\delta)}{2N}} \leq \frac{C_\lambda}{\sqrt{N}} + M\sqrt{\frac{\log(2/\delta)}{2N}}.
\]

\emph{Third term.} The function $f^\star$ does not depend on the sample, so its losses are independent and lie in $[0,M]$. The $i$th has mean $\E_{P_{t_i}}[\ell(f^\star)]$, so their average has mean $\sum_j \frac{B_j}{N}\E_{P_j}[\ell(f^\star)] = \E_Q[\ell(f^\star)]$. By Hoeffding's inequality, for $\varepsilon > 0$,
\[
P\bigl(\hat{\E}[\ell(f^\star)] - \E_Q[\ell(f^\star)] \geq \varepsilon\bigr) \leq \exp\Biggl(-\frac{2N^2\varepsilon^2}{\sum_{i=1}^{N} M^2}\Biggr) = \exp\Biggl(-\frac{2N\varepsilon^2}{M^2}\Biggr).
\]
This is the same tail as for $\Phi$, so with probability at least $1-\delta/2$,
\[
\hat{\E}[\ell(f^\star)] - \E_Q[\ell(f^\star)] \leq M\sqrt{\frac{\log(2/\delta)}{2N}}.
\]
Adding the two bounds, and using $2M\sqrt{\log(2/\delta)/(2N)} = M\sqrt{2\log(2/\delta)/N}$, gives the lemma.
\end{proof}

\section{Proof of Lemma~\ref{lem:fold}}\label{app:fold}

\begin{proof}[Proof of Lemma~\ref{lem:fold}]
Given $\mathcal{D}_{\mathrm{train},k}$, the fit $\hat{f}_{\lambda,k}$ is a fixed function. Training and validation times in fold $k$ are disjoint, so by Assumption~2 each validation observation at time $t$ is independent of the fit and has distribution $P_t$. By linearity, with $B_t$ observations at each $t \in V_k$,
\begin{align*}
\E\bigl[\widehat{\mathrm{CV}}_k(\lambda) \bigm| \mathcal{D}_{\mathrm{train},k}\bigr]
&= \frac{1}{N_{\mathrm{val},k}}\sum_{t\in V_k}\sum_{i=1}^{B_t}\E\bigl[\ell(\hat{f}_{\lambda,k}(X_{t,i}),\, Y_{t,i}) \bigm| \mathcal{D}_{\mathrm{train},k}\bigr] \\
&= \sum_{t\in V_k}\frac{B_t}{N_{\mathrm{val},k}}\,\E_{P_t}[\ell(\hat{f}_{\lambda,k})]
= \mathrm{CV}_k(\lambda).
\end{align*}
Each gap compares two models scored under different distributions. $\mathcal{R}(\lambda)$ scores $f^\star_{\lambda,Q_h}$ under $P_{T+s}$, and $\mathrm{CV}_k(\lambda)$ scores $\hat{f}_{\lambda,k}$ under the validation periods. We can bound how much worse one model is than the other under a training mixture. Under $Q_h$, $f^\star_{\lambda,Q_h}$ is best in $\mathcal{F}_\lambda$, so it is no worse than the fit. Under $Q^{(k)}$, by Lemma~\ref{lem:erm}, the fit is worse than the best in $\mathcal{F}_\lambda$ by at most an estimation term. So each case moves both scores to one such mixture, compares the models there, and pays for each move. A move changes the distribution under which a loss in $[0,M]$ is averaged (Assumption~4), so by the bounded-loss inequality it costs at most $M$ times a total variation distance, which Lemma~\ref{lem:drift} bounds. The weights $\pi_t$ sum to one, so it suffices to bound each validation period $t$ and average.
\begin{itemize}
    \item \emph{Underestimate.}
    Fix $t \in V_k$ and compare the two models under $Q_h$:
    \begin{align*}
    \E_{P_{T+s}}[\ell(f^\star_{\lambda,Q_h})] - \E_{P_t}[\ell(\hat{f}_{\lambda,k})]
    &= \bigl(\E_{P_{T+s}}[\ell(f^\star_{\lambda,Q_h})] - \E_{Q_h}[\ell(f^\star_{\lambda,Q_h})]\bigr) \\
    &\quad + \bigl(\E_{Q_h}[\ell(f^\star_{\lambda,Q_h})] - \E_{Q_h}[\ell(\hat{f}_{\lambda,k})]\bigr) \\
    &\quad + \bigl(\E_{Q_h}[\ell(\hat{f}_{\lambda,k})] - \E_{P_t}[\ell(\hat{f}_{\lambda,k})]\bigr).
    \end{align*}
    The middle summand is at most zero, since $\hat{f}_{\lambda,k} \in \mathcal{F}_\lambda$ and $f^\star_{\lambda,Q_h}$ minimizes the risk under $Q_h$. Lemma~\ref{lem:drift} with the weights $\gamma$ bounds the first summand by $Mq$ and the last by $M\sum_{j'}\gamma_{j'}\bigl(L|t-j'|^\alpha \wedge 1\bigr)$. Averaging over $t$,
    \[
    \mathcal{R}(\lambda)-\mathrm{CV}_k(\lambda) \leq Mq + Mu_k.
    \]
    Both risks in the gap lie in $[0,M]$, so the gap is also at most $M$.
    \item \emph{Overestimate.}
    Fix $t \in V_k$ and compare the two models under $Q^{(k)}$:
    \begin{align*}
    \E_{P_t}[\ell(\hat{f}_{\lambda,k})] - \E_{P_{T+s}}[\ell(f^\star_{\lambda,Q_h})]
    &= \bigl(\E_{P_t}[\ell(\hat{f}_{\lambda,k})] - \E_{Q^{(k)}}[\ell(\hat{f}_{\lambda,k})]\bigr) \\
    &\quad + \bigl(\E_{Q^{(k)}}[\ell(\hat{f}_{\lambda,k})] - \E_{Q^{(k)}}[\ell(f^\star_{\lambda,Q^{(k)}})]\bigr) \\
    &\quad + \bigl(\E_{Q^{(k)}}[\ell(f^\star_{\lambda,Q^{(k)}})] - \E_{Q^{(k)}}[\ell(f^\star_{\lambda,Q_h})]\bigr) \\
    &\quad + \bigl(\E_{Q^{(k)}}[\ell(f^\star_{\lambda,Q_h})] - \E_{P_{T+s}}[\ell(f^\star_{\lambda,Q_h})]\bigr).
    \end{align*}
    Lemma~\ref{lem:drift} with the weights $\omega$ bounds the first summand by $M\sum_{j}\omega_j\bigl(L|t-j|^\alpha \wedge 1\bigr)$ and the last by $Mr_k$. The third summand is at most zero, since $f^\star_{\lambda,Q_h} \in \mathcal{F}_\lambda$ and $f^\star_{\lambda,Q^{(k)}}$ minimizes the risk under $Q^{(k)}$. The second summand is the excess risk of the fit under $Q^{(k)}$. The training block holds $N_{\mathrm{train},k}$ independent observations with mixture $Q^{(k)}$ (Assumption~2), so by Lemma~\ref{lem:erm}, with probability at least $1-\delta$, it is at most $C_\lambda/\sqrt{N_{\mathrm{train},k}} + M\sqrt{2\log(2/\delta)/N_{\mathrm{train},k}}$. This bound does not depend on $t$. On that event, averaging over $t$,
    \[
    \mathrm{CV}_k(\lambda)-\mathcal{R}(\lambda)
    \leq Mb_k + Mr_k + \frac{C_\lambda}{\sqrt{N_{\mathrm{train},k}}} + M\sqrt{\frac{2\log(2/\delta)}{N_{\mathrm{train},k}}}.
    \]
    Both risks in the gap lie in $[0,M]$, so the gap is also at most $M$.
\end{itemize}
\end{proof}

\section{Proof of Theorem~\ref{thm:upper_bound} (Equal-Weight Bound)}\label{app:thm1}

\begin{proof}
The only property of $\hat\lambda$ is that it minimizes $\widehat{\mathrm{CV}}$ over $\Lambda$, so we compare the two risks through the score. Adding and subtracting $\widehat{\mathrm{CV}}(\hat\lambda)$ and $\widehat{\mathrm{CV}}(\lambda^\star)$, the middle term is at most zero since $\lambda^\star \in \Lambda$:
\begin{align*}
    \mathcal{R}(\hat{\lambda})-\mathcal{R}(\lambda^\star)
    &= \bigl(\mathcal{R}(\hat{\lambda}) - \widehat{\mathrm{CV}}(\hat{\lambda})\bigr)
    + \bigl(\widehat{\mathrm{CV}}(\hat{\lambda}) - \widehat{\mathrm{CV}}(\lambda^\star)\bigr)
    + \bigl(\widehat{\mathrm{CV}}(\lambda^\star) - \mathcal{R}(\lambda^\star)\bigr) \\
    &\leq \bigl(\mathcal{R}(\hat{\lambda}) - \widehat{\mathrm{CV}}(\hat{\lambda})\bigr)
    + \bigl(\widehat{\mathrm{CV}}(\lambda^\star) - \mathcal{R}(\lambda^\star)\bigr).
\end{align*}
Call the two remaining terms $(A)$ and $(B)$. Here $(A)$ is how far the score underestimates the risk of the selected configuration, and $(B)$ is how far it overestimates the risk of the oracle. Each mixes validation noise with the drift and fitting error of the fold fits, and the two need different tools. To separate them, let
\[
\mathrm{CV}(\lambda) = \frac{1}{K}\sum_{k=1}^{K}\mathrm{CV}_k(\lambda)
\]
be the average of the fold means of Lemma~\ref{lem:fold}. Adding and subtracting it at $\hat\lambda$ and at $\lambda^\star$,
\begin{align*}
    (A) &= \underbrace{\bigl(\mathcal{R}(\hat{\lambda}) - \mathrm{CV}(\hat{\lambda})\bigr)}_{(A_1)} + \underbrace{\bigl(\mathrm{CV}(\hat{\lambda}) - \widehat{\mathrm{CV}}(\hat{\lambda})\bigr)}_{(A_2)}, \\
    (B) &= \underbrace{\bigl(\widehat{\mathrm{CV}}(\lambda^\star) - \mathrm{CV}(\lambda^\star)\bigr)}_{(B_1)} + \underbrace{\bigl(\mathrm{CV}(\lambda^\star) - \mathcal{R}(\lambda^\star)\bigr)}_{(B_2)}.
\end{align*}
The pieces $(A_2)$ and $(B_1)$ are validation noise, and $(A_1)$ and $(B_2)$ are drift and fitting error. The excess risk is at most $(A_1)+(A_2)+(B_1)+(B_2)$.

    \begin{itemize}
    \item \emph{$(A_1)$.}
    The underestimate bound of Lemma~\ref{lem:fold} holds for every $\lambda$ and every draw of the training blocks. Averaging over the $K$ folds,
    \[
    \mathcal{R}(\lambda)-\mathrm{CV}(\lambda)
    \leq \frac{1}{K}\sum_{k=1}^{K}\min\bigl(M,\; Mq + Mu_k\bigr).
    \]
    The right-hand side depends on neither $\lambda$ nor the sample, so it bounds $(A_1)$ even though $\hat\lambda$ depends on the sample.

    \item \emph{$(A_2)$ and $(B_1)$.}
    Write $D_k(\lambda) = \widehat{\mathrm{CV}}_k(\lambda) - \mathrm{CV}_k(\lambda)$ for the deviation of fold $k$. Then
    \[
    (A_2) = \frac{1}{K}\sum_{k=1}^{K} \bigl(-D_k(\hat\lambda)\bigr),
    \qquad
    (B_1) = \frac{1}{K}\sum_{k=1}^{K} D_k(\lambda^\star).
    \]
    Fix $\lambda \in \Lambda$.

    \begin{itemize}
    \item \emph{Martingale differences.} Folds that overlap have dependent deviations, so we sum the deviations observation by observation. Write $\ell_{k,t,i} = \ell(\hat{f}_{\lambda,k}(X_{t,i}),\, Y_{t,i})$ and $\mu_{k,t} = \E_{P_t}[\ell(\hat{f}_{\lambda,k})]$, so that $\mathrm{CV}_k(\lambda) = \sum_{t \in V_k} (B_t/N_{\mathrm{val},k})\,\mu_{k,t}$ and
    \[
    D_k(\lambda) = \frac{1}{N_{\mathrm{val},k}}\sum_{t \in V_k}\sum_{i=1}^{B_t}\bigl(\ell_{k,t,i} - \mu_{k,t}\bigr).
    \]
    Let $K_t = \{k : t \in V_k\}$ be the folds that validate on period $t$. Since $V_k$ is $m$ consecutive periods, $K_t$ has at most $m$ folds. Exchanging the order of summation,
    \[
    \frac{1}{K}\sum_{k=1}^{K} D_k(\lambda) = \sum_{t=1}^{T}\sum_{i=1}^{B_t}\xi_{t,i},
    \qquad
    \xi_{t,i} = \frac{1}{K}\sum_{k \in K_t}\frac{\ell_{k,t,i} - \mu_{k,t}}{N_{\mathrm{val},k}},
    \]
    with $\xi_{t,i} = 0$ when $K_t$ is empty. Order the observations by period, and within a period by $i$. Each fold $k \in K_t$ trains on periods $k,\ldots,k+h-1$, all before $t$, so $\xi_{t,i}$ is a function of the observations up to $(X_{t,i}, Y_{t,i})$, and given the observations before it, the fits $\hat{f}_{\lambda,k}$ and means $\mu_{k,t}$ are fixed. By Assumption~2, $(X_{t,i}, Y_{t,i})$ is independent of the observations before it, so $\xi_{t,i}$ has conditional mean zero, and the $\xi_{t,i}$ form a martingale difference sequence. Since $\ell_{k,t,i} \in [0,M]$ (Assumption~4), $\xi_{t,i}$ lies between $-\frac{1}{K}\sum_{k \in K_t}\mu_{k,t}/N_{\mathrm{val},k}$, a function of the earlier periods, and that value plus
    \[
    c_t = \frac{M}{K}\sum_{k \in K_t}\frac{1}{N_{\mathrm{val},k}}.
    \]

    \item \emph{Azuma's inequality.} Period $t$ contributes $B_t$ terms with range $c_t$, so by the general form of Azuma's inequality \citep{mcdiarmid1989method}, for either sign and every $\varepsilon > 0$,
    \[
    P\Bigl(\pm\frac{1}{K}\sum_{k=1}^{K} D_k(\lambda) \geq \varepsilon\Bigr) \leq \exp\Bigl(-\frac{2\varepsilon^2}{V}\Bigr),
    \qquad
    V = \sum_{t=1}^{T} B_tc_t^2.
    \]
    By Cauchy--Schwarz over the at most $m$ folds in $K_t$, $c_t^2 \leq (M^2m/K^2)\sum_{k \in K_t} N_{\mathrm{val},k}^{-2}$. Fold $k$ lies in $K_t$ exactly for $t \in V_k$, and $\sum_{t \in V_k} B_t = N_{\mathrm{val},k}$, so with $N_{\mathrm{val},\min} = \min_k N_{\mathrm{val},k}$,
    \[
    V \leq \frac{M^2m}{K^2}\sum_{k=1}^{K}\sum_{t \in V_k}\frac{B_t}{N_{\mathrm{val},k}^2} = \frac{M^2m}{K^2}\sum_{k=1}^{K}\frac{1}{N_{\mathrm{val},k}} \leq \frac{M^2m}{KN_{\mathrm{val},\min}}.
    \]
    Setting $\exp(-2\varepsilon^2KN_{\mathrm{val},\min}/(M^2m))$ to $\delta'$ and solving, with probability at least $1-\delta'$,
    \[
    \pm\frac{1}{K}\sum_{k=1}^{K} D_k(\lambda) \leq M\sqrt{\frac{m\log(1/\delta')}{2KN_{\mathrm{val},\min}}}.
    \]

    \item \emph{The bound on $(A_2)$.} The configuration $\hat\lambda$ depends on the sample, so a bound for one fixed $\lambda$ does not apply at $\hat\lambda$. We therefore take the minus sign at every $\lambda \in \Lambda$, $J$ events, each with $\delta' = \delta/(3J)$. By the union bound they all hold with probability at least $1-\delta/3$, and then the last display holds at every $\lambda$, in particular at $\hat\lambda$:
    \[
    (A_2) \leq M\sqrt{\frac{m\log(3J/\delta)}{2KN_{\mathrm{val},\min}}}.
    \]

    \item \emph{The bound on $(B_1)$.} The configuration $\lambda^\star$ does not depend on the sample, so we take the plus sign at $\lambda^\star$ alone, with $\delta' = \delta/3$. With probability at least $1-\delta/3$,
    \[
    (B_1) \leq M\sqrt{\frac{m\log(3/\delta)}{2KN_{\mathrm{val},\min}}}.
    \]

    \end{itemize}

    \item \emph{$(B_2)$.}
    The configuration $\lambda^\star$ does not depend on the sample. Apply the overestimate bound of Lemma~\ref{lem:fold} at $\lambda^\star$ on each fold with failure probability $\delta/(3K)$, so that $\log(2/\delta)$ becomes $\log(6K/\delta)$. A union bound over the $K$ folds and $C_{\lambda^\star} \leq C_{\max}$ give, with probability at least $1-\delta/3$,
    \[
    (B_2) \leq \frac{1}{K}\sum_{k=1}^{K}\min\Biggl(M,\; Mb_k + Mr_k + \frac{C_{\max}}{\sqrt{N_{\mathrm{train},k}}} + M\sqrt{\frac{2\log(6K/\delta)}{N_{\mathrm{train},k}}}\Biggr).
    \]

    \item \emph{Sum.}
    The bound on $(A_1)$ holds for every draw. The bounds on $(A_2)$, $(B_1)$, and $(B_2)$ each fail with probability at most $\delta/3$, so all hold together with probability at least $1-\delta$. Adding them,
    \begin{align*}
    \mathcal{R}(\hat{\lambda})-\mathcal{R}(\lambda^\star)
    &\leq \underbrace{\frac{1}{K}\sum_{k=1}^{K}\min\bigl(M,\; Mq + Mu_k\bigr)}_{(A_1)} \\
    &\quad + \underbrace{\frac{1}{K}\sum_{k=1}^{K}\min\Biggl(M,\; Mb_k + Mr_k + \frac{C_{\max}}{\sqrt{N_{\mathrm{train},k}}} + M\sqrt{\frac{2\log(6K/\delta)}{N_{\mathrm{train},k}}}\Biggr)}_{(B_2)} \\
    &\quad + \underbrace{M\sqrt{\frac{m\log(3J/\delta)}{2KN_{\mathrm{val},\min}}}}_{(A_2)}
    + \underbrace{M\sqrt{\frac{m\log(3/\delta)}{2KN_{\mathrm{val},\min}}}}_{(B_1)}.
    \end{align*}
    This is the bound of Theorem~\ref{thm:upper_bound}.
\end{itemize}
\end{proof}

\section{Proof of Theorem~\ref{thm:weighted_bound} (Weighted Bound)}\label{app:thm2}

\begin{proof}
    The opening of Appendix~\ref{app:thm1} applies with $\widehat{\mathrm{CV}}_w$ in place of $\widehat{\mathrm{CV}}$, and with
    \[
    \mathrm{CV}_w(\lambda) = \sum_{k=1}^{K} w_k\, \mathrm{CV}_k(\lambda)
    \]
    in place of $\mathrm{CV}(\lambda)$. The excess risk is at most the same four pieces.

    \begin{itemize}
    \item \emph{$(A_1)$ and $(B_2)$.}
    The $(A_1)$ and $(B_2)$ steps of Appendix~\ref{app:thm1} go through with $\sum_k w_k$ in place of the average over folds, on the same event for the fits. The weights do not enter the training blocks, so each fold's estimation term is unchanged.

    \item \emph{$(A_2)$ and $(B_1)$.}
    With $D_k(\lambda)$ as in Appendix~\ref{app:thm1}, now
    \[
    (A_2) = \sum_{k=1}^{K} w_k\bigl(-D_k(\hat\lambda)\bigr),
    \qquad
    (B_1) = \sum_{k=1}^{K} w_k D_k(\lambda^\star).
    \]
    Since the weights are fixed, the steps of Appendix~\ref{app:thm1} apply with $w_k$ in place of $1/K$. The range of each $\xi_{t,i}$ becomes $c_t(w) = M\sum_{k \in K_t} w_k/N_{\mathrm{val},k}$, and Cauchy--Schwarz over the at most $m$ folds in $K_t$ and $\sum_{t \in V_k} B_t = N_{\mathrm{val},k}$ give
    \[
    V(w) = \sum_{t=1}^{T} B_tc_t(w)^2 \leq M^2m\sum_{k=1}^{K}\frac{w_k^2}{N_{\mathrm{val},k}} \leq \frac{M^2m\|w\|_2^2}{N_{\mathrm{val},\min}}.
    \]
    As in Appendix~\ref{app:thm1}, the minus sign at every $\lambda \in \Lambda$, each with $\delta' = \delta/(3J)$, gives with probability at least $1-\delta/3$
    \[
    (A_2) \leq M\|w\|_2\sqrt{\frac{m\log(3J/\delta)}{2N_{\mathrm{val},\min}}},
    \]
    and the plus sign at $\lambda^\star$ with $\delta' = \delta/3$ gives with probability at least $1-\delta/3$ the same bound on $(B_1)$ with $\log(3/\delta)$ in place of $\log(3J/\delta)$.

    \item \emph{Sum.}
    As in Appendix~\ref{app:thm1}, the four bounds hold together with probability at least $1-\delta$. Adding them,
    \begin{align*}
    \mathcal{R}(\hat{\lambda})-\mathcal{R}(\lambda^\star)
    &\leq \underbrace{\sum_{k=1}^{K} w_k\min\bigl(M,\; Mq + Mu_k\bigr)}_{(A_1)} \\
    &\quad + \underbrace{\sum_{k=1}^{K} w_k\min\Biggl(M,\; Mb_k + Mr_k + \frac{C_{\max}}{\sqrt{N_{\mathrm{train},k}}} + M\sqrt{\frac{2\log(6K/\delta)}{N_{\mathrm{train},k}}}\Biggr)}_{(B_2)} \\
    &\quad + \underbrace{M\|w\|_2\sqrt{\frac{m\log(3J/\delta)}{2N_{\mathrm{val},\min}}}}_{(A_2)}
    + \underbrace{M\|w\|_2\sqrt{\frac{m\log(3/\delta)}{2N_{\mathrm{val},\min}}}}_{(B_1)}.
    \end{align*}
    This is the bound of Theorem~\ref{thm:weighted_bound}.
\end{itemize}
\end{proof}
\newpage
\section{Bounds Under a General Latent Index}\label{app:general_theta}

The main text specializes to $\theta(t) = t$ (Section~\ref{sec:setup}). Under a general latent map $\theta$, the bounds hold with latent distances in place of calendar gaps.

Lemma~\ref{lem:drift} carries over with $d_\Theta(\theta(t),\, \theta(j))^\alpha$ in place of $|t-j|^\alpha$, by the same proof. With $V_k$, $\pi_t$, $\omega_j$, and $\gamma_{j'}$ as in Lemma~\ref{lem:fold}, and $\Delta(t,j) = L\, d_\Theta\bigl(\theta(t),\, \theta(j)\bigr)^\alpha \wedge 1$, set
\begin{align*}
    q &= \sum_{j'=T-h+1}^{T}\gamma_{j'}\, \Delta(T{+}s,\, j'),
    &
    u_k &= \sum_{t\in V_k}\pi_t\sum_{j'=T-h+1}^{T}\gamma_{j'}\, \Delta(t,\, j'),\\
    r_k &= \sum_{j=k}^{k+h-1}\omega_j\, \Delta(T{+}s,\, j),
    &
    b_k &= \sum_{t\in V_k}\pi_t\sum_{j=k}^{k+h-1}\omega_j\, \Delta(t,\, j).
\end{align*}
The bounds of Theorems~\ref{thm:upper_bound} and~\ref{thm:weighted_bound} then hold with these sums. Under $\theta(t) = t$ the latent distances are the calendar gaps, and the sums are those of Lemma~\ref{lem:fold}.

\section{CPS Application Details}\label{app:cps}

\paragraph{Sample.}
We use the ASEC samples of 1976 to 2026 from IPUMS CPS \citep{ipumscps2026}. The earnings year is the survey year minus one. A household is interviewed in two consecutive March supplements, so we keep households in their first one (months in sample 1 to 4) and drop the ASEC oversample, which keeps each household in a single year. In 2014 we keep the traditional questionnaire (the 5/8 file). A worker is eligible if aged 18 to 64, a wage and salary worker outside the armed forces, at work at least 50 weeks of the year and usually at least 35 hours a week, with positive wage and salary income that is not imputed and a known education. Imputation is read from the earnings allocation flag through the 1987 survey and from the flags for other and longest-job earnings after it. We keep one randomly chosen eligible worker per household, which leaves about $8{,}000$ to $15{,}000$ workers per year.

\paragraph{Outcome and characteristics.}
Annual wage and salary income is converted to 1999 dollars and capped at $\$100{,}000$, below the lowest real top code in any year ($\$101{,}100$, in the 1981 survey), so the cap is the same in every year. It is divided by weeks worked times usual weekly hours, logged, and clipped to $[\log 2,\, \log(100{,}000/1{,}750)]$. Predictions are clipped to the same range. Education has five groups (less than high school, high school, some college, bachelor's degree, graduate degree), harmonized across the 1992 change from years of schooling to degrees. Race and Hispanic origin have four groups (Hispanic, and non-Hispanic white, Black, and other), coarse because of the 2003 change in the race question. Marital status has three groups (married, never married, other). Age enters as a quadratic. Two breaks remain, the 2014 questionnaire redesign and the updated processing system introduced with the 2019 survey.

\paragraph{Assumptions.}
Draws are with replacement, so Assumption~2 holds exactly. Assumption~5 holds with $J = 5$, and clipping gives Assumption~4 with $M = (\log 28.6)^2 \approx 11.2$.

\paragraph{Protocol.}
Before any weighted comparison we fixed the sample, the outcome, the configurations, $h = 5$, $m = 1$, $B_t = 2000$, the $\rho$ grid, and the half-life rule, and checked only that the best configuration within each year's own data changes over the record. It changes once, from sex to education in 1983. The deployment years, first set to 1990 to 2025, were moved to 1981 to 2025 after that check, since deployment years that begin seven years after the change cannot show how quickly each weighting follows it. The smaller $B_t$ of Figure~\ref{fig:cps}(b) and the windows of Table~\ref{tab:cps_h} were added after the first comparison.

\paragraph{Results.}
Under $\mathcal{R}$ the lead of education grows from $0.005$ in 1983 to $0.02$ by 1990, against risks near $0.28$. With $B_t = 2000$, every weighting pays when deploying in 1983, since every fold then validates on a year up to 1982, when sex was still best. Equal weights peak at $0.012$ in 1987, the half-life rule recovers by 1988, and the last fold stays below $0.0025$ after 1983. The folds that favor sex have the largest $u_k$ and $r_k$, and equal weights give them as much weight as the newest folds. Across $B_t$, equal weights lose $0.0007$ to $0.0019$ to the best $\rho$, the last fold loses $0.012$ at $B_t = 25$ and is best at $B_t = 2000$, and the half-life rule is within $0.0002$ of the best for $B_t \leq 250$ and $0.0004$ behind at $B_t = 2000$.

\paragraph{Window width.}
Table~\ref{tab:cps_h} repeats the comparison for $h \in \{1, 3, 5, 10\}$ over the deployment years 1986 to 2025, which every window can reach, with 100 replications. Each $h$ has its own oracle (Section~\ref{sec:pop_min}), so the table compares weights within a row, not windows. At $B_t = 50$ the excess risk is U-shaped in $\rho$ for every $h$, and at $B_t = 2000$ it is smallest at $\rho \leq 0.5$ for $h \leq 5$. What equal weights lose against the best $\rho$ shrinks as $h$ grows, from $0.0083$ at $h = 1$ to $0.0003$ at $h = 10$ for $B_t = 50$, and at $h = 10$ and $B_t = 2000$ every weighting is within $0.0001$ of the best.

\begin{table}[H]
\centering
\caption{Excess deployment risk ($\times 10^{3}$) on CPS earnings, deployment years 1986 to 2025.}
\label{tab:cps_h}
\begin{tabular}{rrccccccc}
\toprule
 & & \multicolumn{7}{c}{$\rho$} \\
\cmidrule(lr){3-9}
$h$ & $B_t$ & $0$ & $0.5$ & $0.7$ & $0.8$ & $2^{-1/5}$ & $0.9$ & $1$ \\
\midrule
1 & 50 & 17.6 & 11.3 & 8.5 & 7.6 & 7.9 & 8.3 & 15.9 \\
1 & 2000 & 0.2 & 0.1 & 0.2 & 0.4 & 0.8 & 1.0 & 2.4 \\
3 & 50 & 12.8 & 5.9 & 3.4 & 2.7 & 2.8 & 3.0 & 4.7 \\
3 & 2000 & 0.2 & 0.0 & 0.1 & 0.2 & 0.5 & 0.6 & 1.5 \\
5 & 50 & 11.8 & 4.6 & 2.6 & 2.1 & 2.2 & 2.3 & 3.4 \\
5 & 2000 & 0.1 & 0.0 & 0.1 & 0.2 & 0.3 & 0.4 & 0.8 \\
10 & 50 & 11.1 & 4.5 & 2.4 & 2.0 & 1.9 & 2.0 & 2.2 \\
10 & 2000 & 0.2 & 0.1 & 0.1 & 0.1 & 0.1 & 0.1 & 0.1 \\
\bottomrule
\end{tabular}
\end{table}

\section{Flight Delay Application Details}\label{app:flights}

\paragraph{Sample.}
We use the monthly Reporting Carrier On-Time Performance files of the Bureau of Transportation Statistics \citep{bts2026ontime} for January 2010 to December 2025, which cover the domestic flights of the reporting carriers. We drop cancelled and diverted flights, which have no arrival delay. This leaves about $400{,}000$ to $650{,}000$ flights per month, and $170{,}000$ to $370{,}000$ from April 2020 to February 2021. Flights share airports, days, and weather, so they are not independent, but draws with replacement from one month's flights are, so Assumption~2 holds exactly for $P_t$ as defined.

\paragraph{Outcome and configurations.}
We clip the arrival delay in minutes to $[-30, 120]$, which gives Assumption~4 with $M = 150^2 = 22{,}500$. The departure hour has the 19 blocks of the scheduled departure time, the carriers are the 23 that report in any month of the record, each airport configuration has the 30 busiest airports over the record and one group for the rest, and the day of the week has seven. Each configuration predicts the mean of $Y$ within each group, and a group with no training flights gets the training mean. Assumption~5 holds with $J = 5$.

\paragraph{Protocol.}
We chose this record after the earnings application, to test changes that do not last. Before any weighted comparison we fixed the sample, the outcome, the configurations, $h = 5$, $m = 1$, $B_t \in \{25, 50, 100, 250, 2000, 10000\}$, the limit in which every fold is scored on all flights of its months, the $\rho$ grid, the half-life rule, and the deployment months July 2010 to December 2025, copying Appendix~\ref{app:cps} where we could. We checked only that the best configuration within each month's own data, fit on half its flights and scored on the other half, changes over the record and changes back. It does. Departure hour is best in most summer months of every year but 2020, and a carrier, airport, or day-of-week configuration in many winter months, by small margins. The calendar-month view of Figure~\ref{fig:flights}(a) was chosen after the comparison.

\paragraph{Drift.}
Under $\mathcal{R}$, departure hour leads by up to $88$ in summer, against risks between $355$ and $1{,}452$. In the 49 months when another configuration is best, its median lead is $3.3$. Of the 21 spells of such months, 18 last one or two months, and the longest, 15 months, spans the pandemic in 2020 and 2021.

\paragraph{Results.}
With $B_t = 2000$, equal weights select departure hour in every deployment month, so they lose only when another configuration is best. The last fold loses to them by $4.2$ in December and gains by $1.1$ in October. From $B_t = 250$ up, the last fold loses to equal weights by $5.6$ at $B_t = 250$, $1.1$ at $2000$, $0.56$ at $10{,}000$, and $0.35$ in the limit without validation noise. The half-life rule is within $0.09$ of the best $\rho$ from $B_t = 2000$ up and $2.0$ behind at $250$. At $B_t \leq 100$ each fold trains on at most $500$ flights, too few for the configurations with 19 to 31 groups, so the seven-group day-of-week model wins the fold scores, and equal weights select it in about 97\% of deployments at $B_t = 25$. $\mathcal{R}$ scores the configurations fit on all flights, where departure hour is best, so this is the estimation term of Lemma~\ref{lem:erm}, and the last fold does best there only because its noise spreads its picks.

\paragraph{Seasonal weights.}
After the comparison we tried weighting folds by calendar month, with equal weight on the folds that validate in the deployment's calendar month, and with weights that halve for each month of seasonal distance. Without validation noise the first has excess risk $1.20$ and the second $1.22$, against $1.22$ for equal weights. With noise the first does worse, $1.47$ at $B_t = 2000$ and $3.84$ at $B_t = 250$ over 40 replications, since it uses one fold in twelve. A configuration other than departure hour that is best in a month is best again in the same month a year later only 12\% of the time.

\end{document}